\documentclass[12pt]{article}
\usepackage{psfig,epsf,amssymb,amsmath,latexsym}
\title{Low density expansion for Lyapunov exponents
}

\author{Hermann Schulz-Baldes
\\
\\
{\small Mathematisches Institut, Universit\"at
Erlangen-N\"urnberg,}
\\
{\small Bismarckstra{\ss}e 1$\frac{1}{2}$,
91054 Erlangen, Germany}
}

\date{ }

\newcommand{\CC}{{\mathbb C}}
\newcommand{\NN}{{\mathbb N}}
\newcommand{\RR}{{\mathbb R}}

\newcommand{\ZZ}{{\mathbb Z}}

\newcommand{\pp}{{\bf p}}
\newcommand{\EE}{{\bf E}}

\newcommand{\Ss}{{\cal S}}
\newcommand{\Oo}{{\cal O}}

\newcommand{\Nn}{{\cal N}}

\begin{document}

\maketitle

\begin{abstract}
In some quasi-one-dimensional weakly disordered media, 
impurities are large and rare rather than small and dense. 
For an Anderson model with a low density of strong impurities, a
perturbation theory  in the impurity density is developed for the
Lyapunov exponent and the density of states. The Lyapunov exponent
grows linearly with the density. Anomalies of the Kappus-Wegner
type appear for all rational quasi-momenta even in lowest order
perturbation theory.
\end{abstract}

\vspace{.5cm}

\section{Introduction}

A perturbative formula for the Lyapunov exponent of a one-dimensional
random medium for weakly coupled disorder was
first given by Thouless \cite{Tho} and then proven
rigorously by Pastur and Figotin \cite{PF}. Anomalies in the
perturbation theory at the band
center were discovered by Kappus and Wegner \cite{KW} and further
discussed by various other authors \cite{DG,BK,CK}. The Lyapunov
exponent is then identified with the inverse localization length of
the system. This short note concerns
the behavior of the Lyapunov exponent for a low density of impurities,
each of which may, however, be large. The presented method is as
\cite{JSS,SB,SSS} a further application of
diagonalizing the transfer matrices without perturbation (here the low
density of impurities) and then rigorously controlling the error terms
by means of oscillatory sums of rotating modified 
Pr\"ufer phases. Some of the
oscillatory sums remain large if the rotation phases (here the
quasi-momenta) are rational. This leads to supplementary contributions
of the Kappus-Wegner type.

\vspace{.2cm} 

The calucalations are carried through
explicitely for the one-dimensional Anderson model, 
but the method transposes also
to more complicated models with a periodic background as well as 
low-density disorder with correlations similar to the random polymer model
\cite{JSS}. Extension to a quasi-one-dimensional situation as in
\cite{SB} should be possible, but is even more cumbersome on a calculatory
level. It is also straightforward to calculate and control higher
order terms in the density.

\vspace{.2cm} 

As one motivation for this study (apart from a mathematical one), let
us indicate that a low density of strong impurities
seems to describe 
materials like carbon nanotubes more adequately than a small coupling
limit of the Anderson model. Indeed, these materials have perfect
cristaline structure over distances of microns which leads to a
ballisistic transport over such a distance \cite{FPWD}. The existing 
few defects are, on the other hand, quite large.
Coherent transport within a one-particle framework should then be
studied by a model similar to the one considered here.
However, it is possible that 
the impurties rather play the role of quantum dots so that
Coulomb blockade is the determining effect for the transport properties 
\cite{MBCYL} rather than the coherent transport studied here.

\section{Model and preliminaries}

The standard one-dimensional Anderson Hamiltonian is
given by

$$
(H_\omega\psi)_n
\;=\;
-\,\psi_{n+1}\,-\,\psi_{n-1}\,+\,v_n\psi_{n}
\;,
\qquad
\psi\in\ell^2(\ZZ)
\;.
$$

\noindent Here $\omega=(v_n)_{n\in\ZZ}$ is a sequence of independent and
identically distributed real random variables. The model is determined
by their probability distribution $\pp$ depending on 
a given density $\rho\in[0,1]$:

\begin{equation}
\label{eq-measure}
\pp
\;=\;
(1-\rho)\,\delta_0+ \rho\,\tilde{\pp}
\;,
\end{equation}

\noindent where $\tilde{\pp}$ is a fixed compactly supported
probability mesure on $\RR$. 
This measure may simply be a Dirac peak 
if there is only one type of impurity, but 
different from $\delta_0$. Set $\Sigma=\mbox{supp}(\pp)$
and $\tilde{\Sigma}=\mbox{supp}(\tilde{\pp})$. The expectation
w.r.t. the $\pp$'s will be denoted by $\EE$, that w.r.t. the  
$\tilde{\pp}$'s by $\tilde{\EE}$, while $\EE_v$ and
$\tilde{\EE}_v$ is the expectation w.r.t. $\pp$ and $\tilde{\pp}$
respectively over one random variable $v\in\Sigma$ only.

\vspace{.2cm}

In order to define the Lyapunov exponent,
one rewrites the Schr\"odinger equation $H_\omega \psi=E\psi$ using
transfer matrices

$$
\left(\begin{array}{c} \psi_{n+1} \\ \psi_{n}
\end{array} \right)
\;=\;
{T}_n^E\;
\left(\begin{array}{c}\psi_n \\ \psi_{n-1}
\end{array} \right)
\mbox{ . }
\qquad
{T}_n^E
\;=\;
\left(\begin{array}{cc} v_n-E & -1 \\ 1 & 0
\end{array} \right)
\;.
$$

\noindent We also write $T^E_v$ for $T^E_n$ if $v_n=v$. 
Then the Lyapunov exponent at energy $E\in\RR$ associated to products
of random matrices chosen independently according to $\pp$ from the
family $(T^E_v)_{v\in\Sigma}$ of SL$(2,\RR)$-matrices 
is given by

\begin{equation}
\label{eq-Lypa0}
\gamma(\rho,E)
\;=\;
\lim_{N\to\infty}\frac{1}{N}\;\EE\,
\log\left(\Bigl\|\prod_{n=1}^N T^E_n
\Bigr\|\right)
\;.
\end{equation}

\noindent The aim is to study the asymptotics of 
$\gamma(\rho,E)$ in small $\rho$ for $|E|<2$. 

\vspace{.2cm}

In order to state our results,
let us introduce, for $E=-2\cos(k)$ with $k\in(0,\pi)$, 
the basis change $M\in\,$SL$(2,\RR)$
and the rotation matrix $R_k$ by the quasi-momentum $k$:

$$
M\;=\;
\frac{1}{\sqrt{\sin(k)}}
\left(\begin{array}{cc} \sin(k) & 0
\\ -\cos(k) & 1 \end{array}\right)
\;,
\qquad
R_k\;=\;
\left(\begin{array}{cc} \cos(k) & -\sin(k)\\ \sin(k) & \cos(k) 
\end{array}\right)
\mbox{ . }
$$

\noindent It is then a matter of computation to verify

$$
MT^E_{v}M^{-1}
\;=\;
R_k({\bf 1}+\,P^E_v)
\mbox{ , }
\qquad
P^E_v\;=\;
-\,\frac{v}{\sin(k)}
\left(\begin{array}{cc} 0 & 0
\\ 1 & 0 \end{array}\right)
\mbox{ . }
$$
 
Next we introduce another auxillary family of random matrices. Set
$\hat{\Sigma}=[-\frac{\pi}{2},\frac{\pi}{2})\times\tilde{\Sigma}$ and,
for $(\psi,v)\in\hat{\Sigma}$:

$$
\hat{T}^E_{\psi,v}
\;=\;
R_\psi\,
MT^E_v M^{-1}
\;.
$$

\noindent The following probability measures on $\hat{\Sigma}$ will
play a role in the sequel: 
$\hat{\pp}_\infty=\frac{d\psi}{\pi}\otimes \tilde{\pp}$ and
$\hat{\pp}_q=\left(\frac{1}{q}\sum_{p=1}^q
\delta_{\frac{\pi}{2}(\frac{p}{q}-\frac{q+1}{2q})}\right)\otimes \tilde{\pp}$
for $q\in\NN$. The Lyapunov exponents associated to 
these families of random matrices are
denoted by $\hat{\gamma}_\infty(E)$ and $\hat{\gamma}_q(E)$
respectively. It is elementary to check that the subgroups generated by
matrices corresponding to
the supports of $\hat{\pp}_\infty$ and $\hat{\pp}_q$ are non-compact,
which implies \cite{BL} that the corresponding 
Lyapunov exponents are strictly positive.

\vspace{.2cm}

The matrices $T^E_{v}$ and $\hat{T}^E_{\psi,v}$ induce actions
$\Ss_{E,v}$ and $\hat{\Ss}_{E,\psi,v}$ on $\RR$ via

\begin{equation}
e_{\Ss_{E,v}(\theta)}
\;=\;
\frac{MT^E_{v}M^{-1}e_{\theta}}{
\|MT^E_{v}M^{-1}e_{\theta}\|}
\mbox{ , }
\qquad
e_{\hat{\Ss}_{E,\psi,v}(\theta)}
\;=\;
\frac{\hat{T}^E_{\psi,v}e_{\theta}}{
\|\hat{T}^E_{\psi,v}e_{\theta}\|}
\mbox{ . }
\label{eq-action}
\end{equation}

\noindent where the freedom of phase is fixed by 
$\Ss_{E,0}(\theta)=\theta+k$ and
$\hat{\Ss}_{E,\psi,0}(\theta)=\theta+k+\psi$ as well as the continuity in
$v$. Invariant measures $\nu^E$, $\hat{\nu}^E_\infty$ and 
$\hat{\nu}^E_q$ for these actions and the probability measures  
$\pp$, $\hat{\pp}_\infty$ and $\hat{\pp}_q$ are defined by

$$
\int^\pi_0 d\nu^E(\theta)\,f(\theta)
\;=\;
\int^\pi_0 d\nu^E(\theta)\,\EE_v\,f(\Ss_{E,v}(\theta)\mbox{mod} \pi)
\mbox{ , }
\qquad
f\in C(\RR/ \pi\ZZ)
\mbox{ , }
$$

\noindent and similar formulas for $\hat{\nu}^E_\infty$ and 
$\hat{\nu}^E_q$.
Due to a theorem of Furstenberg \cite{BL}, the invariant measures 
exist and are unique
whenever the associated Lyapunov exponent is positive.
Let us note that one can easily verify that the invariant measure
$\hat{\nu}^E_\infty$
is simply given by the Lebesgue measure $\frac{d\theta}{\pi}$.
Furthermore $\hat{\nu}^E_\infty$ and $\hat{\nu}^E_q$ do not depend on
$\rho$ (but $\nu^E$ does).

\vspace{.2cm}

Next let us write out a
more explicit formula for the new Lyapunov exponent
$\hat{\gamma}_\infty(E)$. First of all,
according to Furstenberg's formula \cite{BL,JSS}, 

$$
\hat{\gamma}_\infty(E)
\;=\;
\hat{\EE}_{\psi,v}\;
\int^\pi_0  d\hat{\nu}^E_\infty(\theta)\;
\log
(\|\hat{T}^E_{\psi,v}\, e_\theta\|)
\;,
\qquad
\mbox{where }\;
e_\theta
\;=\;
\left(\begin{array}{c} \cos(\theta) \\ \sin(\theta) 
\end{array}\right)
\;.
$$

\noindent As $\hat{\nu}^E_\infty$ is the Lebesgue measure,
rotations are orthogonal and the integrand is $\pi$-periodic, one gets

\begin{equation}
\label{eq-Lyap1}
\hat{\gamma}_\infty(E)
\;=\;
\frac{1}{2}\;\tilde{\EE}_{v}\;
\int^{2\pi}_0  \frac{d\theta}{2\pi}\;
\log
\Bigl(\langle e_\theta|
(MT^E_v M^{-1})^*(MT^E_{v}M^{-1})|e_\theta\rangle\Bigr)
\;.
\end{equation}

\noindent Now $(MT^E_v M^{-1})^*(MT^E_{v}M^{-1})=|{\bf
1}+P^E_v|^2$ is a
positive matrix with eigenvalues $\lambda_v\geq 1$ and
$1/\lambda_v$ where
$\lambda_v=1+\frac{a}{2}+\sqrt{a+\frac{a^2}{4}}$ with
$a=\frac{v^2}{\sin^2(k)}$ . As it can be diagonalized by an orthogonal
transformation leaving the Lebesgue measure invariant, we deduce that

$$
\hat{\gamma}_\infty(E) 
\;=\;
\frac{1}{2}\;
\tilde{\EE}_{v}
\int^{2\pi}_0  \frac{d\theta}{2\pi}\;
\log
\Bigl(\lambda_v\cos^2(\theta)
+\frac{1}{\lambda_v}\sin^2(\theta)\Bigr)
\;=\;
\frac{1}{2}\;
\int d\tilde{\pp}(v) 
\;
\log
\Bigl(
\frac{1+\lambda^2_v}{2\lambda_v}
\Bigr)
\;.
$$

\noindent This formula shows immediately that
$\hat{\gamma}_\infty(E)>0$ unless $\tilde{\pp}=\delta_0$ (in which
case $\lambda_v=\lambda_0=1$).

\section{Result on the Lyapunov exponent}

\noindent {\bf Theorem}
{\it
Let $E=-2\cos(k)$ with either $\frac{k}{\pi}$ rational or $k$
satisfying the weak diophantine condition

\begin{equation}
\label{eq-Dio}
\bigl|1-e^{2\imath m k}\bigr|
\;\geq\;
c\,e^{-\xi'|m|}
\;,
\qquad
\forall\;\;m\in\ZZ
\;,
\end{equation}

\noindent for some $c>0$ and $\xi'>0$. Then

$$
\gamma(\rho,E)
\;=\;
\left\{
\begin{array}{ccc}
\rho\,\hat{\gamma}_\infty(E)\,+\,\Oo(\rho^2)
& & k\;\mbox{\rm satisfies (\ref{eq-Dio}) },
\\
& & \\
\rho\,\hat{\gamma}_q(E)\,+\,\Oo(\rho^2) 
& &
k=\pi\,\frac{p}{q}\;, 
\end{array}
\right.
$$

\noindent where $p$ and $q$ are relatively prime. Furthermore,  for $\xi$
depending only on $\,\tilde{\pp}$,

$$
\left|\,
\hat{\gamma}_q(E)-
\hat{\gamma}_\infty(E)
\,
\right|
\;\leq \;
c\,e^{-\xi|q|}
\;.
$$

}

\vspace{.3cm}

The result can be interpreted as follows: if the density of the
impurities is small, then the incoming (Pr\"ufer) phase at the
impurity is uniformly distrubuted for a sufficiently irrational rotation
angle ({\it i.e.} quasi-momentum) because the sole invariant measure of
an irrational rotation is the Lebesgue measure. For a rational
rotation, the mixing is to lowest order in $\rho$ perfect over the
orbits of the rational rotation, which leads to the definition of the family 
$(\hat{T}^E_{p,\sigma})_{(p,\sigma)\in\hat{\Sigma}_q}$ and
its distribution $\hat{\pp}_q$. As indicated above, the proof that
this is the correct image is another simple application of modified 
Pr\"ufer phases and an oscillatory sum argument.

\vspace{.2cm}

Let us note that $\hat{\gamma}_q(E)\neq \hat{\gamma}_\infty(E)$; more
detailed formulas for the difference are given below.
As a result, one can expect a numerical curve of the energy dependence 
of the Lyapunov exponent at a given fixed low density
to have spikes at energies corresponding to rational
quasimomenta with small denominators. Moreover, the invariant measures
$\nu^E$ and $\hat{\nu}^E_q$ are {\it not} close to the Lebesgue
measure, but has higher harmonics as is typical at Kappus-Wegner
anomalies. Furthermore,
let us add that at the band center $E=0$ the identity
$\gamma(\rho,0)=\rho\,\hat{\gamma}_0(0)$ holds with no higher order 
correction terms and where $\hat{\gamma}_0(0)$ is the center of band
Lyapunov exponent of the usual Anderson model with distribution
$\tilde{\pp}$. 

\vspace{.2cm}

Finally, let us compare the above result with that obtained for a
weak-coupling limit of the Anderson model \cite{PF,JSS}: the Lyapunov
exponent grows quadratically in the coupling constant of the
disordered potential, while it grows linearly in the density. The
reason is easily understood if one thinks of the change of the
coupling constant also rather as a change of the probability
distribution on the space of matrices. At zero coupling, the measure
is supported on one {\it critical} point (or more generally, on a
commuting subset), and the weight in its neighborhood grows as the
coupling constant grows. In (\ref{eq-measure}) the weight may grow far
from the critical point, and this leads to the faster increase of the
Lyapunov exponent.

\section{Proof}

For fixed energy $E$, configuration $(v_n)_{n\in\NN}$ and 
$(\psi_n)_{n\in\NN}$, as well as an initial condition $\theta_0$, 
let us define iteratively the seqences

\begin{equation}
\label{eq-dyn}
\theta_{n}\;=\;
\Ss_{E,v_n}(\theta_{n-1})
\mbox{ , }
\qquad
\hat{\theta}_{n}\;=\;
\hat{\Ss}_{E,\psi_n,v_n}(\hat{\theta}_{n-1})
\;=\;
{\Ss}_{E,v_n}(\hat{\theta}_{n-1})+\psi_n
\mbox{ . }
\end{equation}

\noindent When considered modulo $\pi$, these are also called the
modified Pr\"ufer phases. They can be efficiently used in order 
to calucate the Lyapunov
exponent as well as the density of states. For the Lyapunov exponent,
let us first note that one can make a basis change in (\ref{eq-Lypa0})
at the price of boundary terms vanishing at the limit, and
furthermore, that according to \cite[A.III.3.4]{BL} 
it is possible to introduce an arbitrary initial vector, so that

\begin{equation}
\label{eq-lyap2}
\gamma(\rho,E)
\;=\;
\lim_{N\to\infty}
\frac{1}{N}\,\EE\,\log\left(\Bigl\|
\Bigl(\prod_{n=1}^N \,MT^E_{n}M^{-1}\Bigr)e_\theta\Bigr\|\right)
\mbox{ . }
\end{equation}

\noindent Now using the modified Pr\"ufer phases with initial
condition $\theta_0=\theta$, this can be developed into a telescopic
sum:

\begin{eqnarray}
\gamma(\rho,E)
& = &
\lim_{N\to\infty}
\frac{1}{N}\,\EE\,\sum_{n=1}^N\;
\log\left(\Bigl\|
\,MT^E_{n}M^{-1}e_{\theta_{n-1}}
\Bigr\|\right)
\nonumber
\\
& = &
\rho\;
\lim_{N\to\infty}
\frac{1}{N}\,\EE\,\sum_{n=1}^N\;
\tilde{\EE}_v\;
\log\left(\Bigl\|
\,MT^E_{v}M^{-1}e_{\theta_{n-1}}
\Bigr\|\right)
\;,
\nonumber
\end{eqnarray}

\noindent where in the second step we have evaluated the partial
expectation over the last random variable $v_n$ by using the fact that
for $v_n=0$ the contribution vanishes. Next let us note that the
function $e^{\imath\theta}\mapsto
\tilde{\EE}_v\,\log(\|MT^E_{v}M^{-1}e_{\theta}\|)$ has an analytic
extension to $\CC\backslash\{0\}$, contains only even frequencies so
that its Fourier series 

$$
\tilde{\EE}_v\;
\log\left(\Bigl\|
\,MT^E_{v}M^{-1}e_{\theta}
\Bigr\|\right)
\;=\;
\sum_{m\in\ZZ}
\;
a_m\;e^{2\imath m\theta}
\;,
$$

\noindent has coefficients satisfying for any $\xi>0$ a Cauchy
estimate of the form

\begin{equation}
\label{eq-Cauchy}
|a_m|
\;\leq\;
c_\xi\,e^{-\xi|m|}
\;.
\end{equation}

\noindent Comparing with (\ref{eq-Lyap1}), we deduce

$$
a_0
\;=\;
\hat{\gamma}_\infty(E)\;.
$$

\noindent Introducing now the oscillatory sums

$$
I_m(N)
\;=\;
\EE\;
\frac{1}{N}\,\sum_{n=1}^N\;
e^{2\imath m\theta_n}
\;,
\qquad
\hat{I}_m(N)
\;=\;
\hat{\EE}\;
\frac{1}{N}\,\sum_{n=1}^N\;
e^{2\imath m\hat{\theta}_n}
\;,
$$

\noindent the Lyapunov exponent now reads

\begin{equation}
\label{eq-Lyap3}
\gamma(\rho,E)
\;=\;
\rho\;\sum_{m\in\ZZ}
\;
a_m\;\lim_{N\to\infty}\;I_m(N)
\;.
\end{equation}

\noindent Hence we need to calculate $I_m(N)$ perturbatively in
$\rho$. Clearly $I_0(N)=1$. Furthermore, integrating over the initial
condition w.r.t. the invariant measure gives for all $N\in\NN$

$$
\int d\nu^E(\theta)\;I_m(N)
\;=\;
\int d\nu^E(\theta)\,e^{2\imath m\theta}
\;.
$$

\noindent Hence calculating $I_m(N)$ perturbatively also gives the
harmonics of $\nu^E$ perturbatively (similar statements hold for 
$\hat{I}_m(N)$, of course).
Going back in history once, one gets

\begin{eqnarray}
I_m(N)
& = &
\frac{1}{N}\;\EE\,
\sum_{n=1}^N\;
\Bigl((1-\rho)\,e^{2\imath mk}\,e^{2\imath m\theta_{n-1}}\,
+\,\rho\;\tilde{\EE}_v( e^{2\imath m \Ss_{E,v}(\theta_{n-1})})
\Bigr)
\nonumber
\\
& = & 
(1-\rho)\;e^{2\imath mk}\,I_m(N)
\,+\,
\Oo(\rho,{N}^{-1})
\;.
\nonumber
\end{eqnarray}

\noindent For $k$ satisfying (\ref{eq-Dio}), one deduces

$$
|I_m(N)|
\;\leq \;
\frac{1}{|1-(1-\rho)\;e^{2\imath mk}|}
\;\Oo(\rho,{N}^{-1})
\;\leq\;
c\,e^{\xi'|m|}\;\Oo(\rho,{N}^{-1})
\;.
$$

\noindent Replacing this and (\ref{eq-Cauchy}) with $\xi>\xi'$ into
(\ref{eq-Lyap3}) concludes the proof in this case because only the
term $m=0$ gives a contribution to order $\rho$.

\vspace{.2cm}

If now $k=\pi\frac{p}{q}$, the same argument implies

\begin{equation}
\label{eq-estim}
I_{nq+r}(N)\;=\;
\Oo(\rho,{N}^{-1})
\;,
\qquad
\forall\;\;n\in\ZZ\;,\;\;r=1,\ldots,q-1
\;.
\end{equation}

\noindent Setting

$$
\tilde{\EE}_v(e^{2\imath m\Ss_{E,v}(\theta)})
\;=\;
\sum_{l\in\ZZ}\,b^{(m)}_l\,e^{2\imath (m+l)\theta}
\;,
\qquad
\hat{\EE}_{\psi,v}(e^{2\imath m\hat{\Ss}_{E,\psi,v}(\theta)})
\;=\;
\sum_{l\in\ZZ}\,\hat{b}^{(m)}_l\,e^{2\imath (m+l)\theta}
\;,
$$

\noindent one deduces for the remaining cases

$$
I_{nq}(N)
\;=\;
(1-\rho)\,
I_{nq}(N)
\,+\,
\Oo(N^{-1})\,+\,
\rho\,
\sum_{l\in\ZZ}
\,b^{(nq)}_l\,
\bigl(I_{nq+l}(N)+\Oo(N^{-1})\bigr)
\;.
$$

\noindent Due to (\ref{eq-estim}), this gives the following equations

$$
I_{nq}(N)
\;=\;
\sum_{r\in\ZZ}
b^{(nq)}_{rq}\,
I_{(n+r)q}(N)
\;+\;\Oo((\rho N)^{-1},\rho)
\;.
$$

\noindent They determine the invariant measure $\nu^E$ to lowest order
in $\rho$. This shows, in particular, that $\nu^E$ is already to
lowest order not given by the Lebesgue measure.
We will not solve these equations, but rather show that the
oscillatory sum $\hat{I}_{nq}(N)$ satisfy the same equations, and
hence, up to
higher order corrections, the measure $\hat{\nu}^E_q$ can be 
used instead of $\nu^E$ in order to calculate the Lyapunov
exponent. Indeed, it follows directly from (\ref{eq-dyn}) and the
definition of $\hat{\pp}_q$ that

$$
\hat{b}^{(m)}_l
\;=\;
\delta_{m\!\!\!\!\!\mod\! q=0}\;b^{(m)}_l
\;.
$$

\noindent In particular, $\hat{I}_{m}(N)=0$ if $m\,\mbox{mod} \,q\neq 0$.
Thus we deduce

$$
\hat{I}_{nq}(N)
\;=\;
\sum_{r\in\ZZ}\;
\hat{b}^{(nq)}_{rq}\,
\hat{I}_{(n+r)q}(N)
\;+\;\Oo(N^{-1})
\;=\;
\sum_{r\in\ZZ}\;
b^{(nq)}_{rq}\,
\hat{I}_{(n+r)q}(N)
\;+\;\Oo(N^{-1})
\;.
$$

\noindent Comparing the equations for ${I}_{nq}(N)$ and
$\hat{I}_{nq}(N)$ (which have a unique solution becaue the invariant
measures are unique by Furstenberg's theorem), it follows that 

$$
\hat{I}_{nq}(N)
\;=\;
{I}_{nq}(N)
\;+\;\Oo(\rho,(\rho N)^{-1})
\;.
$$

\noindent Replacing this into (\ref{eq-Lyap3}), one deduces

\begin{eqnarray*}
\gamma(\rho,E)
& = &
\rho\,
\sum_{m\in\ZZ}
\;a_m\,\int d\hat{\nu}^E_q(\theta)\,e^{2\imath m \theta}
\;+\;
\Oo(\rho^2)
\\
& = &
\rho\;\int d\hat{\nu}^E_q(\theta)\;
\tilde{\EE}_v\;
\log\left(\Bigl\|
\,MT^E_{v}M^{-1}e_{\theta}
\Bigr\|\right)
\;+\;
\Oo(\rho^2)
\;.
\end{eqnarray*}

\noindent Now due to the orthogonality of rotations
one may replace $MT^E_{v}M^{-1}$ by $\hat{T}^E_{\psi,v}$, and then the
r.h.s. contains exactly the Furstenberg formula for
$\hat{\gamma}_q(E)$ as claimed. The estimate comparing
$\hat{\gamma}_q(E)$ and $\hat{\gamma}_\infty(E)$
follows directly from the Cauchy estimate (\ref{eq-Cauchy}).

\section{Result on the density of states}

Another ergodic quantity of
interest is the integrated density of states, defined by

$$
\Nn(\rho,E)
\,=\,
\lim_{N\to\infty}
\frac{1}{N}
\,\EE\;\#\,
\Bigl\{ \mbox{negative eigenvalues  
of the restriction of } H_\omega-E \mbox{ to }
\ell^2(\{1,\ldots,N\}) 
\Bigr\}
.
$$

\noindent Recall, in particular, that $\Nn(0,E)=k$ if $E=-2\cos(k)$.
Defining the mean phase shift of the impurities by 

$$
\tilde{\varphi}(\theta)
\;=\;
\tilde{\EE}_v(\Ss_{E,v}(\theta)-\theta)
\;,
$$

\noindent the low density expansion of the density of states 
reads as follows:

$$
\Nn(\rho,E)
\;=\;
\left\{
\begin{array}{ccc}
(1-\rho)\,k\,+\,\rho
\int\frac{d\theta}{2\pi}\,\tilde{\varphi}(\theta)
\,+\,\Oo(\rho^2)
& & k\;\mbox{\rm satisfies (\ref{eq-Dio}) },
\\
& & \\
(1-\rho)\,k\,+\,\rho
\int d\hat{\nu}^E_q(\theta)\,\tilde{\varphi}(\theta)
\,+\,\Oo(\rho^2)
& &
k=\pi\,\frac{p}{q}\;
\end{array}
\right.
$$

\noindent with $p$ and $q$ relatively prime.
The proof of this is completely analogous to the above when
the rotation number calucation ({\it e.g.} \cite{JSS} for a proof) is
applied:

\begin{eqnarray*}
\Nn(\rho,E)
& = &
\lim_{N\to\infty}
\,\frac{1}{N}\,
\EE\,
\sum_{n=1}^N \;
\bigl(\Ss_{E,v}(\theta_{n-1})-\theta_{n-1}\bigr)
\\
& = &
(1-\rho)\,k
\;+\;
\rho\;
\lim_{N\to\infty}
\,\frac{1}{N}\,\EE\,
\sum_{n=1}^N 
\tilde{\varphi}(\theta_{n-1})
\;.
\end{eqnarray*}


\end{document}